# Berry-phase switch in electrostatically-confined topological surface states

Jun Zhang[1], Ye-ping Jiang[1*], Xu-Cun Ma[2,3], and Qi-Kun Xue[2,3,4]

*1 Key Laboratory of Polar Materials and Devices (MOE) and Department of Electronics, East China Normal University, Shanghai 200241, China*

*2 State Key Laboratory of Low-Dimensional Quantum Physics, Department of Physics, Tsinghua University, Beijing 100084, China*

*3 Frontier Science Center for Quantum Information, Beijing 100084, China*

*4 Southern University of Science and Technology, Shenzhen 518055, China*

Here we visualize the trapping of topological surface states in the circular n-p junctions on the top surface of the 7-quintuple-layer three dimensional (3D) topological insulator (TI) $Sb_2Te_3$ epitaxial films. As shown by spatially- and field-dependent tunneling spectra, these trapped resonances show field-induced splittings between the degenerate time-reversal-symmetric states at zero magnetic field. These behaviors are attributed unambiguously to Berry-phase switch by comparing the experimental data with both numerical and semi-classical simulations. The successful electrostatic trapping of topological surface states in epitaxial films and the observation of Berry-phase switch provide a rich platform of exploiting new ideas for TI-based quantum devices.

* Corresponding authors. Email: ypjiang@clpm.ecnu.edu.cn

In quantum mechanics, Berry phase [1] is a fundamental concept that describes the geometric property of electronic bands of a solid, manifesting itself by performing closed motion in the momentum space and showing anomaly when the motion encloses the singularity of Berry curvature. This singularity lies in the anomaly of the geometrical property of Bloch bands [2], such as band crossing or twist of bands with different parities. The three-dimensional (3D) topological insulators (TI) in the Bi(Sb)Te(Se) family [3-7] possess a single band crossing (Dirac point) in the surface state structure, acting as a source of Berry curvature and leading to a Berry phase of $\pi$ for the surface states [8,9]. For confined motion of two-dimensional (2D) Dirac electrons, the existence of a non-trivial Berry phase will lead to the unique Berry-phase switch behavior [9,10], where a weak magnetic field can induce an abrupt change in the Berry-phase and thus quantized energy of specific trapped state. Nonetheless, due to Klein tunneling, the 2D massless Dirac fermions (DFs) can't be confined completely unless a mass term is introduced [11,12]. Experimentally, the efficient electrostatic trapping of massless DFs can be realized in circular n-p or p-n junctions in Graphene [10,13-20], where the local inverted region is achieved by sub-surface charged defects after tuning the graphene close to the charge-neutrality condition.

For 3D TIs however, this remains challenging because of the heavy native doping in the approximate bulk states and the relatively small bulk gap (0.1~0.2 eV) in these materials. In this work, by controllable growth of high quality ultra-thin epitaxial 3D TI $Sb_2Te_3$ films [5,21], circular n-p junctions were successfully obtained on the top surface to realize efficient electrostatic-trapping of topological surface states. The trapped resonate states are visualized by spatially-resolved scanning tunneling spectroscopy (STS) and show field-induced Berry-phase switch. Our experimental data as well as the numerical and semi-classical calculations supports that this dynamic behavior is unambiguously attributed to the non-trivial Berry phase of the top surface states of this ultra-thin 3D TI film.

By using a combined system of molecular beam epitaxy (MBE) and scanning tunneling microscope (STM), high quality $Sb_2Te_3$ films with $E_D \sim 0$ (Dirac energy being close to the Fermi level) were grown on the $SrTiO_3$ (111) substrates and investigated in-situ by the low temperature STM (Fig. 1(a)). The top surface states of the 7 quintuple-layer (QL) films are only slightly p-doped as a result of the native p-doping and the n-doping from the substrate [21]. The Fermi level in this case lies in the bulk gap to enable the efficient local gating of the top surface states by charged defects (insert), thus inducing local inverted regions with formation of n-doped puddles of various sizes on the p-doped background. The spatially dependent STS (STS mapping) taken at -4.7 meV (insert) reveals a nearly circular-shaped structure with reduced LDOS at the center. Note that STS corresponds to the differential conductance (dI/dV or g) of the tunneling junction and reveals the sample's local density of states (LDOS). Point STS (not shown) indicates that the Dirac point is at ~ -20 meV at the center and at ~ 50 meV at 60 nm away from the center. Therefore, a near-circular n-p junction (as sketched in Fig. 1(b)) coming from a group of sub-surface charged defects is clearly evident. In addition, the multiple peaks in the STS taken at various positions (Fig. S1 of [22]) points to the existence of trapped states, implying the efficient electrostatic trapping of massless DFs of the topological surface states in our circular n-p junction.

The spatial profile of the trapped states and their response to the magnetic field can be revealed by taking spatially dependent STS measurements along a line (marked in Fig. 1(a)) at 0 T and 1 T as shown in Fig. 1(c). The energy range is [-100, 100] meV, which approximately corresponds to the bulk gap of the 7-QL $Sb_2Te_3$ film (see STS in Fig. S1 of [22] and those from [5] for comparison). Thus the trapped states in the n-p junction come from topological surface states. The Dirac energy with reduced LDOS, roughly follows the curve of potential energy profile $U(r)$ described below. The Dirac point ranges from ~ -20 meV (center) to ~ 50 meV (edge) as mentioned above.

A close inspection of the resonate states reveals that new states seem to appear at 1 T (hollow dots) between the original ones compared with the states at 0 T. This behavior is more prominent by comparing the point STS at the center (position 1) at 0 T and 1 T (Fig. S1 of [22]), where new states seem to appear approximately at the mid-energy between the original ones. As demonstrated by both the numerical and semi-classical simulations as well as the field-dependent evolution of these trapped states, this is a clear manifestation of the Berry-phase switch behavior.

Firstly, we simulated the resonate states at 0 T and 1 T (shown in Fig. 1(d)) by solving the Dirac equation $\varepsilon\psi(\mathbf{r}) = [v_F \boldsymbol{\sigma} \cdot \mathbf{p} + U(\mathbf{r})]\psi(\mathbf{r})$ numerically as described in the section II of [22]. Here $\mathbf{p} = \mathbf{q} - \mathbf{A}$, where $\mathbf{r}, \mathbf{p}, \mathbf{q}, \mathbf{A}$ are off-center position, the canonical momentum, the kinetic momentum and the vector potential of magnetic field in the form of symmetric gauge, respectively. We simulated the potential profile of the junction by a screened confining potential for electrons (typical for the case of a screened sub-surface point charge) $U(r) = \mu_0 e^{-r^2} + \mu_\infty$, where $\mu_0$ = -134 meV and $\mu_\infty$ = 105 meV to give the best approximation of the experimental potential profile in the region of $r \leq R_0$ (60 nm). By comparing the experimental data and simulation, resonate states with different indexes can be identified. Here $n$, $m$ are radial and azimuthal quantum numbers, respectively. The states (0, $m$) follow a dashed line shown in Fig. 1(d) [see Fig. S3 of [22] for the identification of other states].

By inspection of the symmetry of these states and comparison with the simulation (see Fig. S4 of [22] for simulated partial contribution from the states with $m = \pm 1/2$), the new states appearing at the center at 1 T are attributed to the ($n$, 1/2) states (hollow dots) that split from energies of the degenerate ($n$, $\pm 1/2$) states. Here we notice that, compared with states with larger $m$, only the split ($n$, 1/2) states are clearly visible due to their exclusively dominant distribution at the center. The degenerate resonances at 0 T and the split ones at 1 T are visualized from the STS maps taken at specific energies (see Fig. S1 of [22]). In this work, we mainly focus on the $m = \pm 1/2$ states.

Secondly, in addition to the nearly perfect match between the experimental data and numerical simulation shown above, the splitting and related anomalous behaviors can be understood more explicitly by the following semi-classical approach.

In the circular n-p junction, the Berry phase $\varphi_B$ of cyclotron motion of trapped Dirac electrons on the top surface depends on the winding of the vector $\mathbf{q}$ or spin around the Dirac point, where a winding number of 1 corresponds to a Berry phase of $\pi$ (due to the spinor nature of the surface states' wave function). For Dirac electrons in presence of magnetic field and absence of electrostatic potential, the winding number is always 1.

The resulting Berry phase of $\pi$ shifts the energy ladder of Landau orbits by half of the energy spaces between adjacent levels, leading to the appearance of zeroth Landau level both in graphene and topological insulators [5,23-25], as well as the half-integer quantum Hall effect in graphene [26].

In the presence of a rotational symmetric potential in the n-p circular junction, the classic orbit of electrons normally precesses, yielding a two-valued momentum field $q(r)$ between the two classical return points $r_1$ and $r_2$ of the orbit (see the two vectors at one position): $q_r = p_r = \pm\sqrt{(\varepsilon - \mu)^2 - (m/r - Br/2)^2}$, $q_\theta = p_\theta - A_\theta = m/r - Br/2$. Following Einstein's argument on the problem of quantizing chaos [27], this two-valued $q(r)$ can then be mapped onto a 2-torus as a single-valued vector field to allow for the application EBK rule, in which the energies of the bounded motion can be considered along two separable coordinates (azimuthal $C_\theta$ and radial $C_R$) by $\oint_{C_\theta} p_\theta r d\theta = 2\pi(n_\theta + \gamma_\theta) - \varphi_B$ (defines $m$) and $\oint_{C_R} p_r dr = 2\pi(n_r + \gamma_R) - \varphi_B$ (determines $n$). In this case, although the Berry phase of the azimuthal motion is always $\pi$ similar to the Landau orbit case, the winding number of $q$ of the additional radial motion can be switched between 0 and 1 by a weak magnetic field (shown below).

Figure 2(a) shows the semi-classical cyclotron orbit for the resonate states (1, $\pm 1/2$) at 0 T and 1 T. Here (1, -1/2) at 0 T is not shown for its being time-reversal-symmetric with the state (1, +1/2). Figure 2(b) and 2(c) are the corresponding vector fields and the winding numbers of $q$ along $C_\theta$ and $C_R$. We see that along $C_\theta$ the winding numbers for (1, $\pm 1/2$) are always 1. In contrast, along $C_R$ the winding numbers for (1, -1/2) and (1, 1/2) behave differently, where the former doesn't change and the latter switches from 0 to 1 at 1 T. This leads to a Berry-phase jump of $\pi$ in the quantization condition along this coordinate for the confined orbital of (1, 1/2), corresponding to a reduction of $\pi$ in the radial action $\oint_{C_R} p_r dr$ and an energy reduction ($\Delta\varepsilon_B$) of about one half of the energy difference ($\Delta\varepsilon$) between (1, -1/2) and (0, -1/2) for the state (1, 1/2). On the contrary, the energy of (1, -1/2) without Berry-phase jump barely changes.

Figure 3 shows the energies of states (1, $\pm 1/2$) at 0 T and 1 T obtained experimentally,

numerically and semi-classically. The energies obtained by this semi-classical approach (see section II of [22] for details) match roughly with those from numerical calculation (Fig. 1(d)). Here the discrepancy lies in the fact that the semi-classical analysis assumes a complete confinement condition ($\gamma_R=1/2$) for the Dirac electrons while in the latter case Klein-tunneling happens.

The critical behavior of Berry-phase switch for states ($n$, +1/2) is captured experimentally by taking the field dependent STS at the center, in perfect agreement with the simulated plot as shown in Fig. 4(a). The field-dependent partial-LDOS (numerical) for $m = -1/2$ and +1/2 in Fig. 4(b) indicate that the Berry-phase switch behavior at the center is exclusively from the $m = \pm 1/2$ states. The Berry-phase switch is signified by the appearances of resonances that are offset approximately by one-half of the energy gaps between adjacent ($n$, -1/2) states at some critical magnetic field $B_c$. Here we note that the peak at ~ 50 meV that appear at 0 T between (2, $\pm 1/2$) and (3, $\pm 1/2$) comes from the (1, $\pm 5/2$) states (see Fig. 1(c)). Because of the non-perfect circular shape of the experimental potential-profile, some ($n$, $|m| > 1/2$) states appear at the 'center' at energies above ~ 50 meV.

In general, for each specific resonance ($n$, $m > 0$) there exists a critical positive field $B_c$ ($m$) at which there is a sudden change in the winding number of the semi-classical orbit along $C_R$. At $B_c$, the electrons at the outer return point $r_2$ satisfy $p_r = 0$ and $p_\theta = 0$, which leads to $B_c = 2m/\ln(\mu_0/(\varepsilon - \mu_\infty))$ (see the dashed curve in Fig. 4 for $B_c(m = 1/2)$ or Fig. S2 of [22] for the curves with different $m$). The switch of the winding number or Berry phase along $C_R$ happens in this case. The critical fields for the occurrence of the split ($n$, +1/2) states roughly follow the dashed curve. In contrast, the resonance ($n$, $m < 0$) only evolves due to the orbital effect in the presence of a positive magnetic field, where $\Delta\varepsilon_{orb}$ results from the $B$ term in the azimuthal action $\oint_{C_\theta} p_\theta r d\theta$ which are opposite for states with opposite $m$. For a critical field $B_c < 1$ T, this large splitting $\Delta\varepsilon_B$ dominates over the orbital splitting $\Delta\varepsilon_{orb}$. In addition, the Zeeman term is ignored here because $\varepsilon_z \sim 10^{-2}$ meV at $B_c$. Note that in the presence of a negative magnetic field, the ($n$, $m < 0$) states, instead of the ($n$, $m > 0$) states, show critical behavior.

In Fig. 4(d) we plot the tunneling conductance at specific energies vs. the magnetic field. We see that there is an increase in the tunneling conductance by 60% at ~ 20 meV and 40% at ~ 38 meV above the critical field due to the split (1, +1/2) and (2, +1/2) states (Fig. 4(c)), respectively. Although the splitting happens abruptly at the critical field, the conductance increases gradually due to the reduced lifetime near the criticality. From the semi-classical point of view, the vertical incident angle of those split ($n$, $m > 0$) states at the outer return point near the critical field (see Fig. S2(b) of [22]) leads to a high transmission through the n-p junction due to Klein tunneling. Here we are aware of the anomaly near the zero energy, where the split (0, +1/2) state appears at a much lower magnetic field (Fig. 4(a) and 4(c)). This may be caused by electron-electron interactions that are not considered in our simulation (the split (0, +1/2) state appear almost at 1 T as shown in Fig. S8 of [22]) and may induce anomaly in the case of a low carrier density in our film.

Topological insulators (TI) have been proved to be highly tunable systems in which many exotic phenomena such as dissipationless quantum anomalous Hall states and error-tolerant Majorana states, etc. can be obtained [28-31]. Nonetheless, the exploitation of potential functionality of the intrinsic properties of TI is unexpectedly rare. The successful trapping of topological surface states in the n-p circular junction on the top surface of a 7-QL 3D TI film allows for the first observation of Berry-phase switch behavior (unique to Dirac fermions) in topological insulators, where large splitting happens abruptly at a weak critical magnetic field between degenerate resonances at zero field. Furthermore, the unambiguous observation of Berry-phase switch behavior in the ultra-thin $Sb_2Te_3$ films indicates the existence of non-trivial Berry phase even near the thickness limit of this 3D TI. Thus, the realization of Berry-phase switch on the epitaxial films indicates that topological surface states provide a rich new platform to exploit switchable optoelectronic applications.

**Acknowledgments**

We thank Y. Hu for helpful discussions. The authors acknowledge the supporting from National Science Foundation of China (Grants No. 61804056, 92065102, 51788104 and 62074092).

**Figure captions**

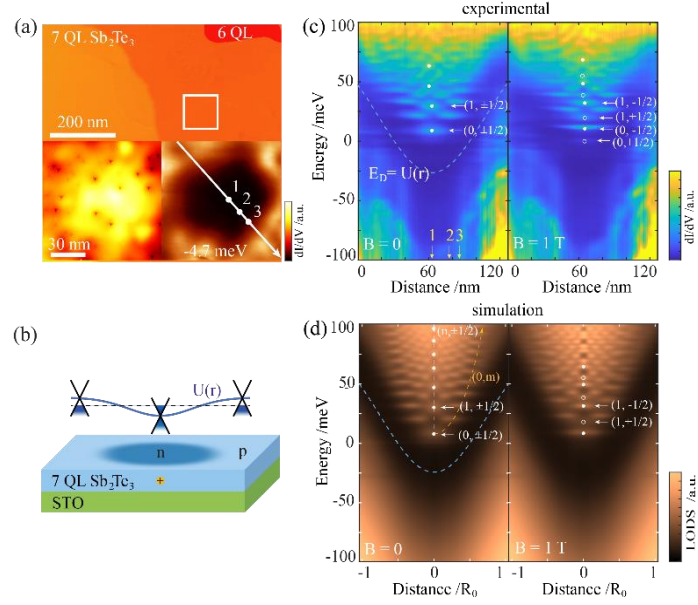

FIG. 1 (color online). (a) Topographic image (tunneling gap: 3 V, 50 pA) of a 7-QL Sb$_2$Te$_3$ film. The insert shows the zoom-in image (left) (1 V, 50 pA) and the STS mapping (right) at - 4.7 meV showing the existence of a circular n-p junction in white-squared region. The point spectra taken at 1, 2 and 3 are shown in Fig. S1 of [22]. (b) Schematic of the circular n-p junction on the top surface of a 7-QL Sb$_2$Te$_3$ film. (c) Spatially resolved STS along the line (126 data points, 125 nm) in (a) at 0 T and at 1 T. These spectra are normalized by *dI/dV* values at 100 meV. The solid dots indicate the ($n$, ±1/2) states at 0 T and ($n$, -1/2) states at 1 T, while the hollow dots indicate the ($n$, 1/2) states at 1 T. The arrows indicate the states (0, ±1/2), (1, ±1/2) at 0 T and at 1 T. (d) Calculated local density of states (LDOS) as function of *r* at 0 T and 1 T for the n-p circular junction of topological surface states. The dashed lines with arrows indicate the resonate states ($n$, ±1/2) and (0, *m*). The dashed curve in (c) and (d) is the potential profile $U(r)$ used in the simulation. All the data were taken at 5.6 K.

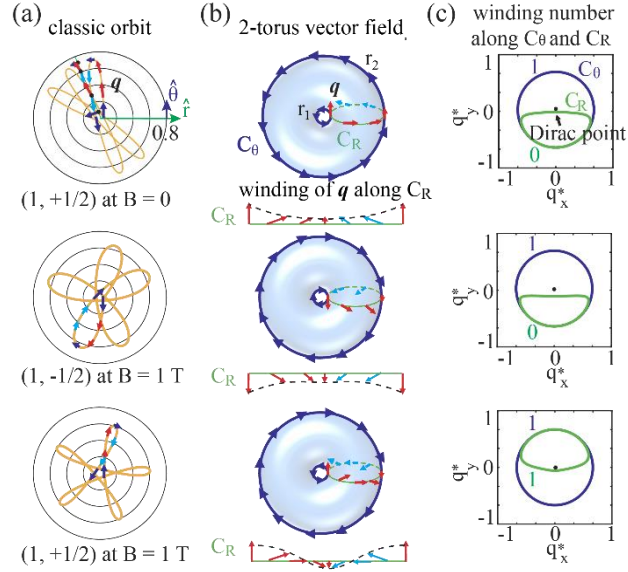

FIG. 2 (color online). (a) The classic orbits of the resonate states (1, ±1/2) at 0 T, (1,-1/2) and (1, +1/2) at 1 T, respectively. The arrows indicate the kinetic momenta $q$. The energies $\varepsilon$ are chosen so that these orbits have radial actions $J_r$ = 1.5, 1.5 and 1.0, respectively [see also section II of [22]]. The red and light blue arrows show that at each position there are two different directions of $q$ due to the precession of electrons. (b) Two-valued vector fields $q(r)$ for the bounded motions in (a) after mapping $q$ onto the 2-torus. Here the arrows indicate the vectors $q$ along two separable coordinates $C_\theta$ and $C_r$. $r_1$ and $r_2$ correspond to the classical turning points of the bounded motion. Here the winding of $q$ along $C_R$ is shown below the vector-field. (c) The corresponding winding of $q^*$ (normalized by $q(r_1)$) with respect to the Dirac point along $C_R$ and $C_\theta$ in the momentum field. The numbers 0 and 1 indicate the winding numbers.

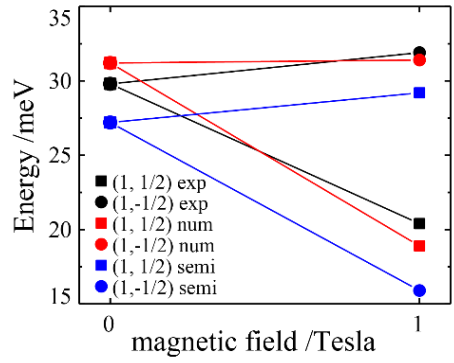

FIG. 3 (color online). Comparison of the energies of (1, ±1/2) states at 0 T and 1T obtained experimentally, numerically and semi-classically.

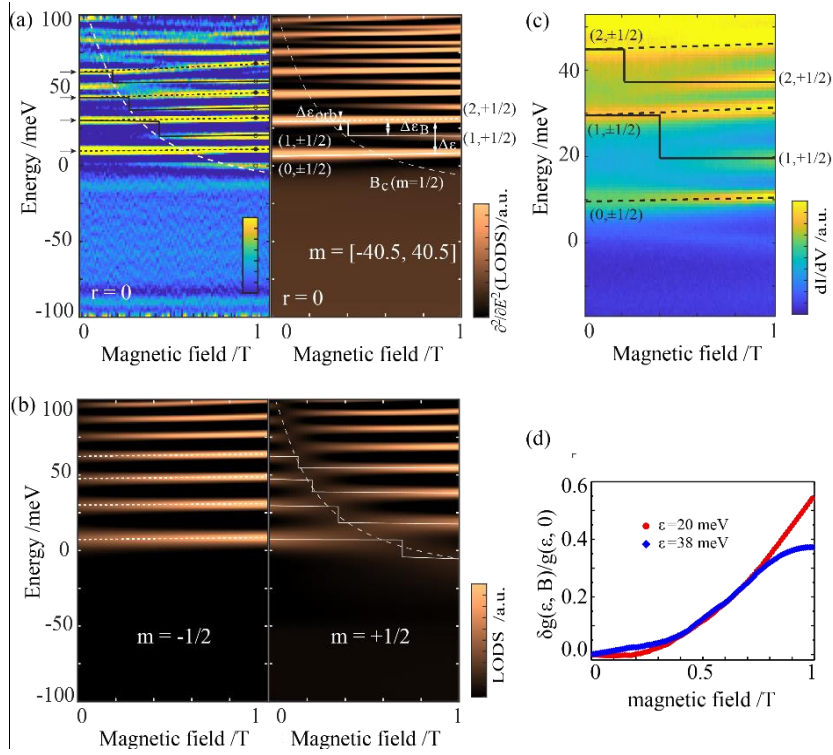

FIG. 4 (color online). (a) Evolutions of LDOS (experimental, 2nd differentiated) and calculated LDOS (numerical, 2nd differentiated) from 0 to 1 T at $r = 0$. The raw data is shown in Fig. S7 of [22]. In the calculation, $m \in [-40.5, 40.5]$. (b) Partial LDOS (numerical) from 0 to 1 T at $r = 0$ for $m = -1/2$ and $+1/2$. The dashed and solid lines indicate the un-split $(n, -1/2)$ and the split $(n, 1/2)$ states, respectively. The dashed curves are the $B_c(\varepsilon)$ plot for $m = 1/2$. $\Delta\varepsilon$, $\Delta\varepsilon_B$ and $\Delta\varepsilon_{orb}$ denote the energy difference between the successive resonance $(n, -1/2)$, the energy jump arising from the Berry-phase switch, and the energy difference between $(n, -1/2)$ and $(n, 1/2)$ below the critical field due to the orbital effect, respectively. (c) LDOS (experimental, raw data) from 0 to 1 T in a smaller energy range. (d) The normalized tunneling conductance (experimental, $\delta g(\varepsilon, B) = (g(\varepsilon, B) - g(\varepsilon, 0))/ g(\varepsilon, 0)$) at 20 meV and 38 meV vs. the magnetic field at the center ($r = 0$). These two energies correspond to approximately the energies of the split $(1, 1/2)$ and $(2, 1/2)$ states shown in (c).

# Supplementary information

# Berry-phase switch in electrostatically-confined topological surface states

Jun Zhang[1], Ye-ping Jiang[1*], Xu-Cun Ma[2,3], and Qi-Kun Xue[2,3,4]

*1 Key Laboratory of Polar Materials and Devices (MOE) and Department of Electronics, East China Normal University, Shanghai 200241, China*

*2 State Key Laboratory of Low-Dimensional Quantum Physics, Department of Physics, Tsinghua University, Beijing 100084, China*

*3 Frontier Science Center for Quantum Information, Beijing 100084, China*

*4 Southern University of Science and Technology, Shenzhen 518055, China*

\* Corresponding authors. Email: ypjiang@clpm.ecnu.edu.cn

**CONTENTS:**

## I. Point STS and mapping of the resonance at specific energies

In choosing the energies at which the STS mapping in Fig. S1(c) and (d) were taken, we referred to the point STS shown in Fig. S1(a) and (b). The positions 1, 2 and 3 are indicated in Fig. 1(a). In Fig. S1, the peaks are carefully indexed according to the data shown in Fig. 1(c).

The degenerate resonances at 0 T and the split ones at 1 T are visualized from the STS maps at specific energies. Figure S1(c) and (d) show spatial LDOS maps of $(0, \pm 1/2)$ and $(1, \pm 1/2)$ states at 0 T and 1 T. The splitting between these states which are degenerate at 0 T is about 10 meV at 1 T, which is nearly one half of energy difference (~ 20 meV) between $(0, \pm 1/2)$ and $(1, \pm 1/2)$ states at 0 T. Furthermore, both the experimental data and numerical simulation show that the energy broadenings of the $(n, m < 0)$ states decrease and the intensities of resonate peaks increase. By contrast, the split $(n, m > 0)$ states behave in the opposite way, with larger energy broadenings and weaker peak intensities. For example, we can see the reduced intensities of the split $(n, +1/2)$ states at 1 T compared with their time-reversal-symmetric $(n, -1/2)$ counterparts, where the intensities of $(0, 1/2)$ and $(1, 1/2)$ are much reduced compared with those of $(0, -1/2)$ and $(1, -1/2)$. This situation is also revealed in the point STS in Fig. S1(b).

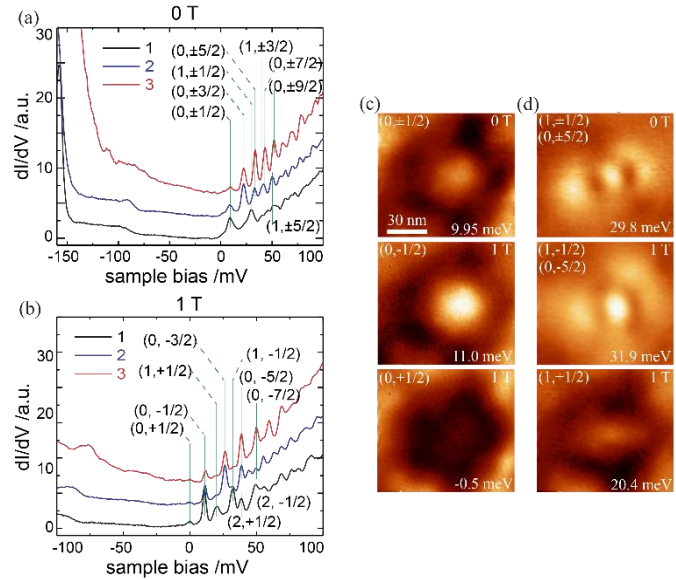

Fig. S1 (a) and (b) Point STS at 0 T and 1 T at positions 1, 2, and 3 in Fig. 1(a). The curves are vertically offset for clarity. The peaks in STS are indexed by referring to both the experimental and simulated data in Fig. 1(c) and 1(d). (c) and (d) STS maps at 0 T and the energy splitting at 1 T for (0, ±1/2) and (1, ±1/2) states, respectively. Some images have two indexes because of the accidental degeneracy.

**II. The simulation of semi-classic orbits in figure 2**

For all the physical terms used in the manuscript, we use the length scale $R_0 = 60$ nm, energy scale $\varepsilon^* = \hbar v_F/R_0 \sim 4.72$ meV. The scale for the magnetic field is thus $B^* = \hbar/(R_0^2 e) \sim 0.183$ T. Then the related equations can be simplified. We use a Fermi velocity $v_F$ of $\sim 4.3\times10^5$ m/s$^2$ according to the literature.

The Dirac Hamiltonian for the surface state electrons in the confining potential defined by $U(r) = \mu_0 e^{-r^2} + \mu_\infty$ is

$$\varepsilon\psi(\boldsymbol{r}) = [v_F \boldsymbol{\sigma} \cdot \boldsymbol{p} + U(\boldsymbol{r})]\psi(\boldsymbol{r}), \tag{S1}$$

where $\boldsymbol{p} = \boldsymbol{q} - \boldsymbol{A}$. Here $p$, $q$, $A$ are chronical momentum, kinetic momentum and the vector potential of magnetic field in the symmetric gauge, respectively. In addition, $\mu_0 = -28.36\varepsilon^*$, $\mu_\infty = 22.23\varepsilon^*$ to give a best fitting to the data in Fig. 2(a) and 2(b). By using the conservation law, $\boldsymbol{p}$ and $\boldsymbol{q}$ are given by

$$q_r = p_r = \pm\sqrt{(\varepsilon - U)^2 - (m/r - Br/2)^2}, \tag{S2}$$

$$q_\theta = p_\theta - A_\theta = m/r - Br/2. \tag{S3}$$

In the central-force situation, the orbit evolves in a precession manner for the bounded electrons. Following Einstein's argument on the problem of quantizing chaos, this circular junction has two constants of motion: the energy and the angular momentum, acting as an integrable system. We immediately see from the above two equations that the orbits define a two-valued momentum field spanned between the classical turning points $r_1$ and $r_2$ (defined by the zeros of $p_r$). The plus and minus signs of $q_r$ define the outward and inward momentum $\boldsymbol{q}$ of the orbit or the two-valued momentum at one point. The two-valued momentum field can then be mapped onto a 2-torus shown in Fig. 2(b).

Here we have a single-valued momentum field. The EBK quantization rule can be readily applied to this integrable system. The energy quantization condition can be considered on two separable coordinates (azimuthal $C_\theta$ and radial $C_R$). The azimuthal

one

$$\oint_{C_\theta} p_\theta r d\theta = 2\pi(n_\theta + \gamma_\theta) - \varphi_B \quad (S4)$$

defines the angular momentum *m*, while the radial one

$$\oint_{C_R} p_r dr = 2\pi(n_R + \gamma_R) - \varphi_B \quad (S5)$$

defines the radial action. The additional phase term $\varphi_B$ defines the Berry phase acquired for electron-motion's winding around the Dirac point (the source of Berry curvature) in the momentum space along $C_\theta$ and $C_R$. Here $\varphi_B/\pi$ defines the winding number. By inspecting the evolution of momenta along $C_\theta$ and $C_R$ in the momentum space, we note that the motion along $C_\theta$ always has a winding number of 1, which leads to the half-integer value of angular momentum number *m*.

In Fig. 2(a), we draw the semi-classical cyclotron orbit at zero magnetic field for the resonate state ($n = 1$, $m = -1/2$) with ($r = r_1$, $\theta = 0$, $p_r = 0$, $p_\theta = -1/2/r_1$) as the initial position. The energy $\varepsilon$ is 5.77 $\varepsilon^*$ to give a radial action $\oint_{C_R} p_r dr = (3/2)2\pi$. Here we take $\gamma_R$ to be 1/2 for the one-dimensional bounded motion $C_R$. The paths of momentum $q^*$ (normalized by $q(r_1)$) along $C_\theta$ and $C_R$ are then plotted in Fig. 2(c), in which we clearly see that the winding number are 1 and 0 along these two coordinates. Here the orbit for ($n = 1$, $m = +1/2$) is not shown for being time-reversal-symmetric with the state ($n = 1$, $m = -1/2$).

In Fig. 2(a), the cyclotron orbits for (1, -1/2) and (1, +1/2) at a magnetic field of 5.5 $B^*$ (~ 1 T) are plotted, with the energies chosen to be 6.20 $\varepsilon^*$ and 3.38 $\varepsilon^*$, respectively. The energy of 6.20 $\varepsilon^*$ keeps the radial action for the state (1, -1/2) at $(3/2)2\pi$, while that of 3.38 $\varepsilon^*$ leads to a radial action of $2\pi$ for the state (1, +1/2). We can see from Fig. 2(c) that the winding number along $C_R$ flips to 1 at some critical field, leading to a Berry phase shift of $\pi$ in equation S5. The additional Berry phase of $\pi$ makes up for the difference of $\pi$ in the radial action. The difference of $2\pi$ in the radial action corresponds to the difference of 1 in the radial quantum number *n*. Thus, if the energy difference between successive (*n*, *m*) states with the same azimuthal number *m* is $\Delta\varepsilon$, the sudden reduction of $\pi$ in the radial quantization (because of the sudden appearance of Berry phase $\pi$) leads to an energy reduction of $\Delta\varepsilon/2$.

The critical field for the resonate states $(n, m)$ to change the winding number along $C_R$ can be found by noting that $p_r$ and $p_\theta$ are both zero at the outer return point in this critical condition. We then get $B_c = 2m/\ln(\mu_0/\varepsilon - \mu_\infty)$, which is plotted in Fig. S2(a). Note that the field of 5.5 $B^*$ is 1 T. Figures S2(b) and S2(c) are the orbit in real space and the corresponding path in the momentum space along $C_R$ for the state (1, 1/2). The energy and the critical field are 3.4 $\varepsilon^*$ and 2.45 $B^*$ (the red point in Fig. S2(a)). We immediately see that the path in the momentum space touches the zero point (Dirac point), indicating the onset of non-trivial winding at this critical field.

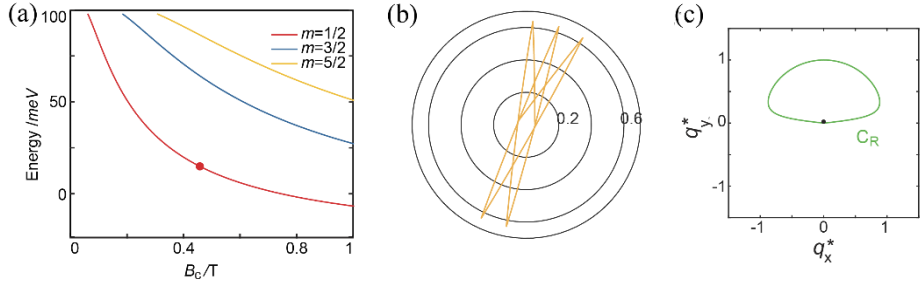

Fig. S2 (a) Plots of the $B_C$-$\varepsilon$ relation for different $m$. (b) The onset of split (1, 1/2) states at the critical field. The energy is 3.4 $\varepsilon^*$ and the critical field is 2.45 $B^*$, corresponding to the red point in s. (c) The path of orbit along $C_R$ in b in the momentum field. 5.5 $B^*$ corresponds to 1 T.

**III. The simulation of resonate states in the n-p Dirac quantum dots**

In the presence of a rotational symmetric field $U(r)$ and by using a symmetric gauge for the uniform vertical magnetic field, for the surface states in the n-p quantum dots the Dirac equation in Eq. S1 can be solved by the radial equation

$$\begin{pmatrix} U(r) - \varepsilon & \partial_r + m/r - Br/2 \\ -\partial_r + m/r - Br/2 & U(r) - \varepsilon \end{pmatrix} \begin{pmatrix} u_1 \\ u_2 \end{pmatrix} = 0 \quad (S6)$$

by using the eigenstates for Eq. S1 in the form of

$$\psi_m(r, \theta) = \frac{e^{im\theta}}{\sqrt{r}} \begin{pmatrix} u_1(r) e^{-i\theta/2} \\ u_2(r) e^{i\theta/2} \end{pmatrix}. \quad (S7)$$

We solve the radial equation S6 by the finite-difference method by using a system size $L = 10$ and the number of sites $N = 2000$. Here we treat the $r = 0$ site by approaching 0 exponentially by adopting a nonlinear grid in the form $\frac{r_i}{STEP} + \frac{\ln r_i}{\ln(RATIO)} + c = i$, where c satisfies $\frac{r_N}{STEP} + \frac{ln r_N}{\ln(RATIO)} + c = N$, $STEP = L/1000$, $RATIO = 1.02$, $r_1 = 0$ and $r_N = L$. Then we have $u_{1\alpha}$, $u_{2\alpha}$ and the corresponding $\varepsilon_\alpha$ by solving the secular equation for the 4000×4000 matrix. In the construction of the matrix for different $m$, we use a simple trick. For $m > 0$, we use the boundary condition of $u_2(0) = 0$, $u_1(L) = 0$. For $m < 0$, we use the boundary condition of $u_1(0) = 0$, $u_2(L) = 0$. This kind of boundary condition preserves the hermiticity of the Hamiltonian and does not change the context of the LDOS map. By using this condition, spurious states are minimized and can be easily removed. These spurious states can be easily identified because they change with different grid parameters.

The local density of states (LDOS) can thus be obtained as the sum of partial contribution of $m$-state contribution $D(\mathcal{E}) = \sum_m D_m(\mathcal{E})$, where

$$D_m(\mathcal{E}) = \sum_\alpha \langle |\psi_{\alpha,m}(r_0,\theta)|^2 \rangle_\lambda \, \delta(\varepsilon - \varepsilon_\alpha). \tag{S8}$$

Here $\alpha$ labels the radial eigenstates of Eq. S6 for fixed $m$, and $\langle |\psi_{\alpha,m}(r,\theta)|^2 \rangle_\lambda$ is a spatial average of the wave function centered at $r_0$. Here the constant $\lambda$ simulates the exponential decay of the LDOS' contribution to the scanning tunneling spectrum with respect to the tunneling center $e^{-(r-r_0)^2/2\lambda^2}$. Then we have

$$D_m(\mathcal{E}) = \sum_\alpha \langle |u_\alpha|^2/r \rangle_\lambda \, \delta(\varepsilon - \varepsilon_\alpha). \tag{S9}$$

Here the delta function is modeled as $\frac{\Gamma}{(\varepsilon - \varepsilon_\alpha)^2 + \Gamma^2}$. We use $\Gamma = 0.5$ and $\lambda = 0.03$ in our simulation, corresponding to an energy broadening of 2.3 meV and a decay length of 2 nm. In the process of plotting the simulation data in figure 1 and figure 4 by summing up all the eigenstates' contribution, the spurious states are excluded. The corresponding ranges of angular momentum [-319.5, 319.5] and [-40.5, 40.5] are used as sufficient conditions to capture all the physical features in the region of our experimental data.

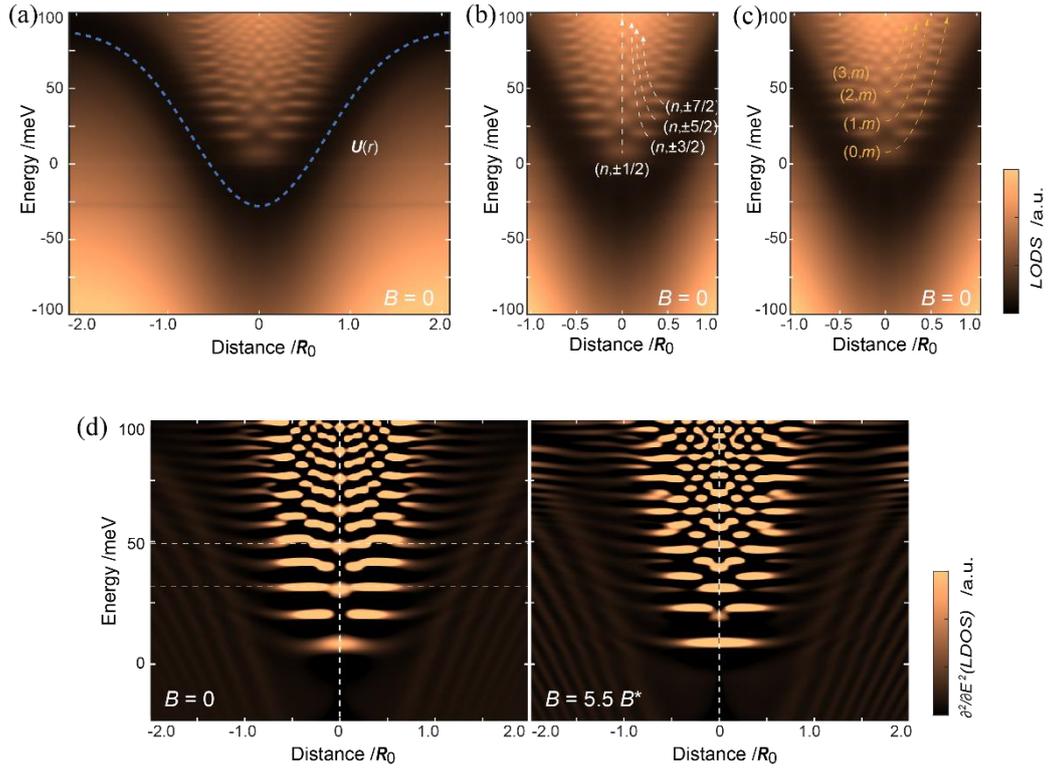

Fig. S3 (a)-(c) Calculated local density of states (LDOS) as function of *r* for an n-p junction of Dirac fermions at $B = 0$ in the energy regions of [-2, +2] and [-1, +1] $R_0$. (d) Comparison of the 2nd differential LDOS at 0 and 5.5 $B^*$ (1 T). The dashed horizontal lines indicate the energies of the $(1, \pm 1/2)$ and the $(2, \pm 1/2)$ states.

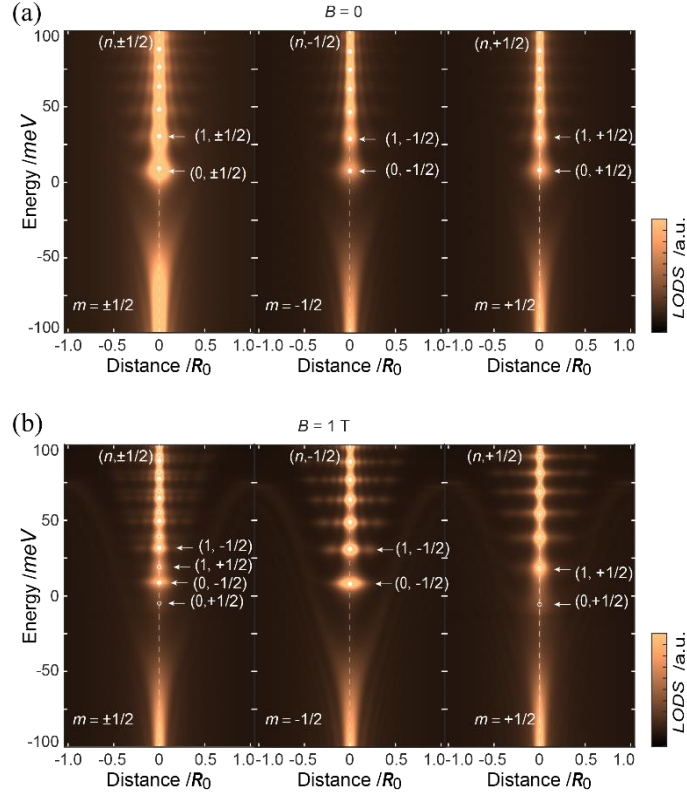

Fig. S4 (a) and (b) Calculated partial LDOS for states ($n$, ±1/2) as function of $r$ for an n-p junction of Dirac fermions at $B = 0$ and 5.5 $B^*$. The degenerate resonances at 0 and their split levels are indicated

## IV. The simulation of field dependent evolution of resonances with specific angular momentum

Figure S5 shows the measured field-dependent LDOS as well as the simulated LDOS at off-center positions 0.24 $R_0$ and 0.39 $R_0$. Different resonances are denoted by different notations. We see that at these two positions, the LDOS is dominated by non-split $m < 0$ states.

To clarify the Berry-phase switch behavior, we simulated separately for the angular momenta $m = +3/2, -3/2$ at $r = 0.24\ R_0$ as shown in Fig. S6, where clear switch behavior can be seen. That the splitting behavior of these states is not clearly visible in the experimental field-dependent data is due to two reasons. First, compared with the levels with negative angular momenta, the split levels with positive angular momenta have decreased intensities because of the reduced lifetimes. Second, the orbital effect on the states with positive and negative angular momenta ($|m| > 1/2$) drives these states in the opposite directions, pushing up the ($m < 0$) states while lowering down the ($m > 0$) states. Thus the split ($n, m > 0$) states that decreased by one-half the energy difference between ($n, m < 0$) and ($n-1, m < 0$) will mix with the ($n-1, m < 0$) at finite magnetic fields. As a result of these two factors, the split states ($|m| > 1/2$) are not clearly visible at off-center positions. For example, in Fig. S6(a) the energy of the split (2, +3/2) state is close to that of the (1, -3/2). In addition, the intensity of this split state is weak compared with the un-split state. We also see that the critical fields for $m = 3/2$ roughly follows the curve $B_c = 2m/\ln(\mu_0/(\varepsilon - \mu_\infty))$.

The critical behavior for states with other angular momentum ($m > 1/2$) can be captured by the simulation at a larger off-center position $r = 0.70\ R_0$ (Fig. S6(d)).

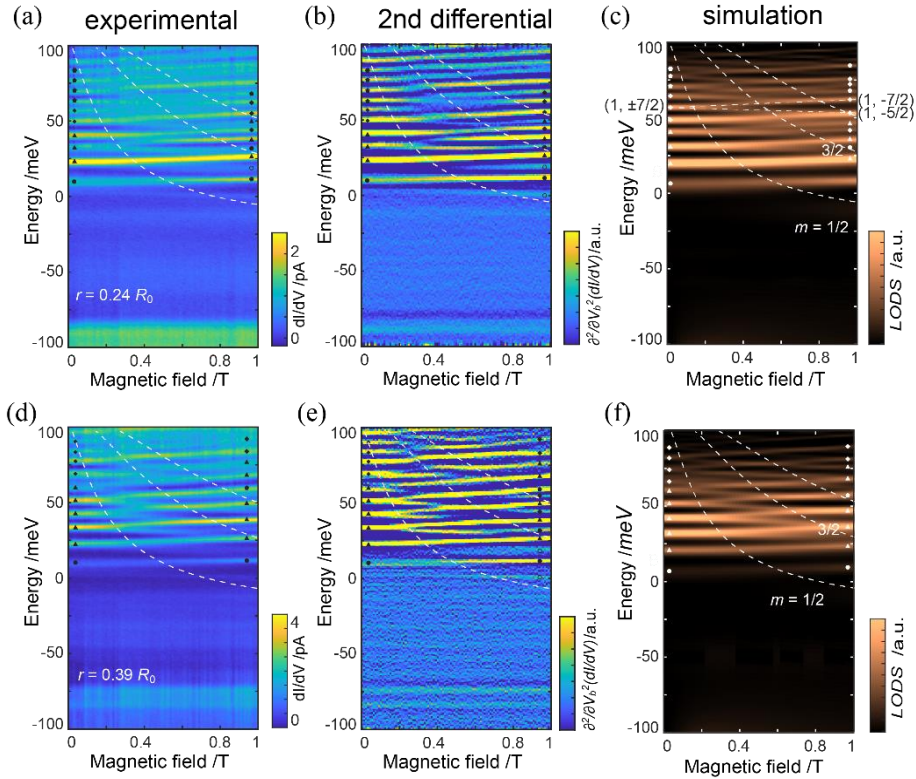

Fig. S5 Background subtracted LDOS, 2nd differential LDOS, and simulated LDOS as function of magnetic fields at r = 0.24 $R_0$ and 0.39 $R_0$, respectively.

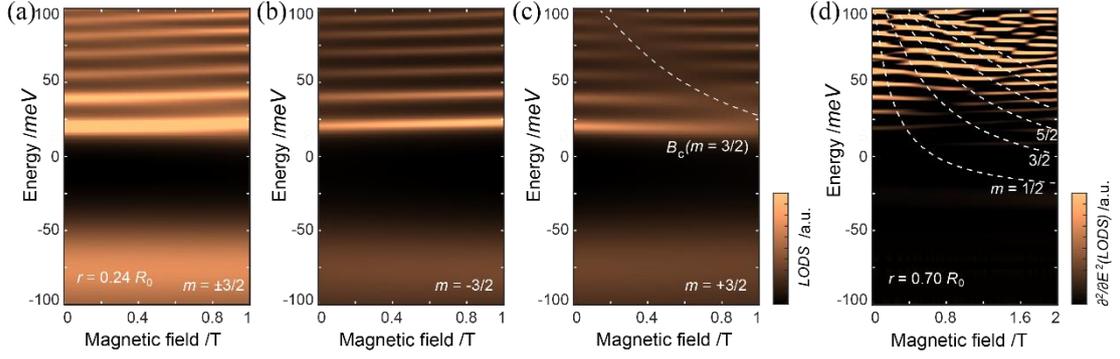

Fig. S6 (a)-(c) Evolutions of partial LDOS of $m = \pm 3/2$ from 0 to 5.5 $B^*$ at $r = 0.24\ R_0$. (d) Calculated LDOS (2nd differentiated) at $r = 0.70\ R_0$.

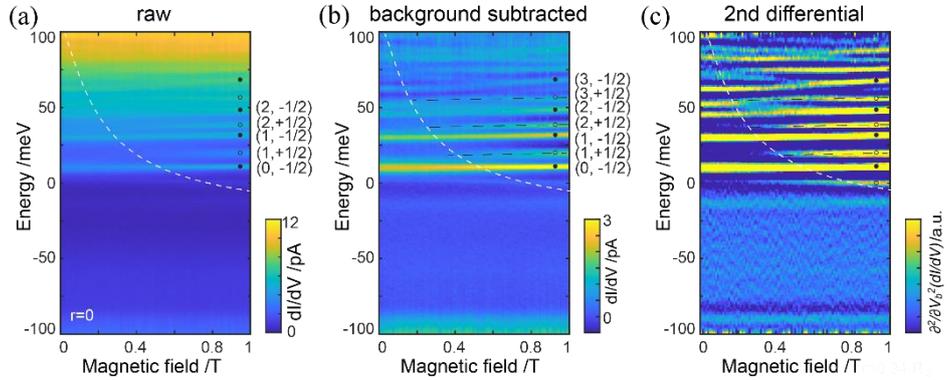

Fig. S7 Raw data, background subtracted data, and 2nd differential data of LDOS as function of magnetic fields from 0 to 1 T at $r = 0$.

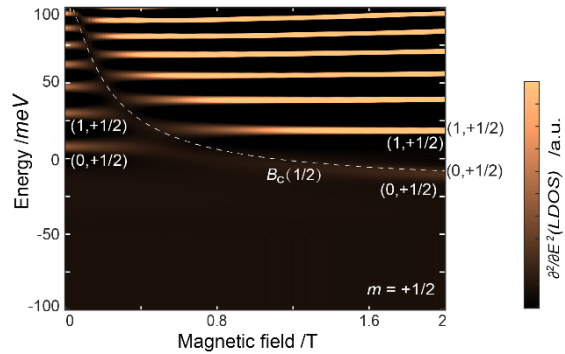

Fig. S8 Field-dependent $(n, +1/2)$ resonances at $r = 0$. Here the field ranges from 0 to 11.0 $B^*$ (2 T).

## V. Sub-surface point charge

Figure S9 shows an STM image where we can clear see the existence of sub-surface point charge. The darker clover-shaped features come from the defect in the topmost quintuple layer (QL). The bright features are from the sub-surface point charges, which are most likely due to the intrinsic defects. This is because we see similar features in the films grown on the graphene substrate, where there is a clean interface between $Sb_2Te_3$ and graphene. The n-type puddles are caused by the intrinsic n-type defects presented in the bottom QL of our films. According to Ref. 21, the n-type defects are Te-on-Sb substitutional defects and appear when Te/Sb flux ratio is high. This is accomplished by a relatively low substrate temperature during the growth of the first QL. During the subsequent deposition of 2-7 QLs the substrate temperature is raised and the resulting defects in the upper layers are all p-type. So the n-type defects only exist in the bottom QL.

The sub-surface point charges can also be introduced intentionally into the bottom QL by extrinsic dopants during the growth of the first QL, or into the STO substrate by the low-dose Argon-sputtering. This enables the large-scale production of the Dirac dots on the surface of topological insulators.

In addition, due to the presence of bulk states, a single sub-surface n-type defect can hardly create an inverted region. So in the current work we focus on a puddle created by a group of n-type defects.

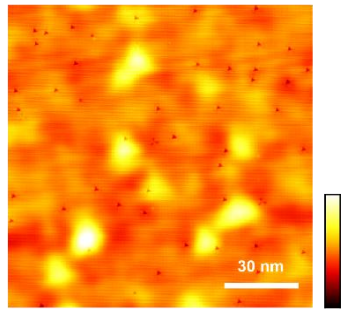

Fig. S9 The STM image (1 V, 50 pA) showing the existence of sub-surface point charge.